# Photo-induced charge transfer across the interface between organic molecular crystals and polymers

V. Podzorov [*] and M. E. Gershenson

*Department of Physics and Astronomy, Rutgers University, Piscataway, New Jersey 08854-8019*

Nov. 09, 2004

Photo-induced charge transfer of positive and negative charges across the interface between an ordered organic semiconductor and a polymeric insulator is observed in the field-effect experiments. Immobilization of the transferred charge in the polymer results in a shift of the field-effect threshold of polaronic conduction along the interface in the semiconductor, which allows for direct measurements of the charge transfer rate. The transfer occurs when the photon energy exceeds the absorption edge of the semiconductor. The direction of the transverse electric field at the interface determines the sign of the transferred charge; the transfer rate is controlled by the field magnitude and light intensity.

*Electronic mail: podzorov@physics.rutgers.edu

Physics of photo-induced charge transfer across organic interfaces, as well as the transport of relaxed polaronic carriers along such interfaces, is very rich and still not well understood. These processes are central in the physics of electronic excitations in organic semiconductors and, therefore, important for applications [1,2]. For example, dissociation of primary photo-excitations (excitons) at the heterojunctions of conjugated polymers leads to an efficient charge transfer and photovoltaic effect [3]. Transport of relaxed excitations (polarons) along the interface between an organic semiconductor and a gate dielectric plays a key role in the organic field-effect transistors (OFETs) [4,5]. Localized charges and polar molecules at the semiconductor/dielectric interface might affect the threshold of the field-induced conductivity [6]. Shift of the field-effect threshold can also be induced in some systems by an application of a strong transverse (gate) electric field - the so-called bias stress effect [7]. Studies of transport and photo-physical properties of interfaces between small-molecule organic semiconductors and polymers are especially important because of the relevance of these systems to all-organic electronic circuits. Better understanding of the light-induced processes at these interfaces is crucial for the success of organic electronics.

In this Letter, we report on observation of the photo-induced charge transfer across the interface between a well-ordered organic molecular semiconductor (rubrene and tetracene) and a polymer (parylene or Mylar). Using the electric field effect at the organic surface as an experimental tool, one can accurately measure the rate of charge transfer across the interface by monitoring the shift of the field-effect threshold. We have observed that, when the interface is illuminated with light of energy $h\nu$ exceeding the absorption edge of the organic semiconductor and an electric field perpendicular to the interface is applied, non-equilibrium charge carriers are injected from the semiconductor into the polymer. The polarity of transferred charges and the transfer rate are controlled by the transverse electric field at the interface, $E_\perp$. The charge injected in the polymer is localized by the deep traps; at room temperature, it does not dissipate for days. This effect results in a reversible shift of the field-effect threshold in OFETs. Potential applications of the observed effect extend from organic light sensors and memory devices to lithography-free patterning of the conduction channel of OFETs by light exposure.

To study the charge transfer between an organic semiconductor and a polymer, we have used the field-effect structures based on organic molecular crystals (OMC) [8,9,10,11]. High quality of the OMC surface facilitates fabrication of reproducible and well-characterized interfaces between OMC and a polymer *parylene* used as the gate dielectric in these devices. Fabrication of the OFETs based on vapor-grown organic molecular crystals and parylene gate dielectric with high mobilities of field-induced charge carriers has been described elsewhere [8]. The source and drain contacts in the studied structures were prepared by depositing the graphite paint on the *a-b* facets of the single crystals [12]; typically, the conduction channel had a width $W \sim 1$ mm and length $L \sim 1$-3 mm. In most of the devices, the gate dielectric was a 1-μm thick parylene film with capacitance per unit area $C_i = 2.3$ nF/cm$^2$ (we also observed the effect reported here with Mylar). The absorption of visible light is negligible in all these polymers. The gate electrode was a thermally evaporated semi-transparent silver film (~ 100 Å thick). The front-gate FET design with a transparent gate dielectric and semi-transparent gate electrode allows studying the light-induced effects at the OMC-polymer interface over a wide spectral range that includes the absorption band of OMC.

All the measurements have been performed in air at room temperature using Keithley source-meters K2400 and electrometers K6512 [13]. The light source in our experiments was a 20-Watts Quartz-Tungsten-Halogen (QTH) lamp with a smooth spectrum in the visible range [14]. To study the spectral response, we have used a set of interference band pass filters ($\Delta\lambda = 20$ nm). The transmission coefficients of the filters and the semi-transparent Ag film (the gate electrode) have been measured using a calibrated silicon light sensor. Unless otherwise specified, the OFETs were illuminated through the semi-transparent gate electrode (see the inset in Fig. 1).

In the dark, the "as-prepared" single-crystal rubrene OFETs, characterized with a high room-temperature mobility of field-induced positive polarons ($\mu = 4$ - 20 cm$^2$/Vs) [5,8], exhibit a very small field-effect threshold (Fig. 1) [15]. This is an indication that the density of deep traps at the OMC-parylene interface does not exceed $10^{10}$ cm$^{-2}$ [8]. The bias stress effect in the rubrene devices is negligible: prolonged application of the gate voltage ($V_g$) in the dark does not affect the dependence of the source-drain current ($I_{SD}$) on the gate voltage $V_g$ (the so-called trans-conductance characteristic of OFETs). This important advantage of rubrene OFETs facilitates the study of photo-induced processes. The central result of this paper is illustrated in Fig. 1: illumination of the OFETs with a visible light while a gate voltage ($V_g^{illum}$) is applied results in a shift of the dark trans-conductance characteristics toward lower or higher $V_g$, depending on the sign of $V_g^{illum}$. The shift is characterized by the magnitude of the field-effect onset voltage ($V_{onset}$) shown by an arrow for one of the curves at the bottom panel of Fig. 1. This effect is observed only if the device is illuminated through a semi-transparent gate electrode; illumination of the rear surface of the crystal does not alter $V_{onset}$, unless the thickness of the crystal is very small ($\leq 100$ μm).

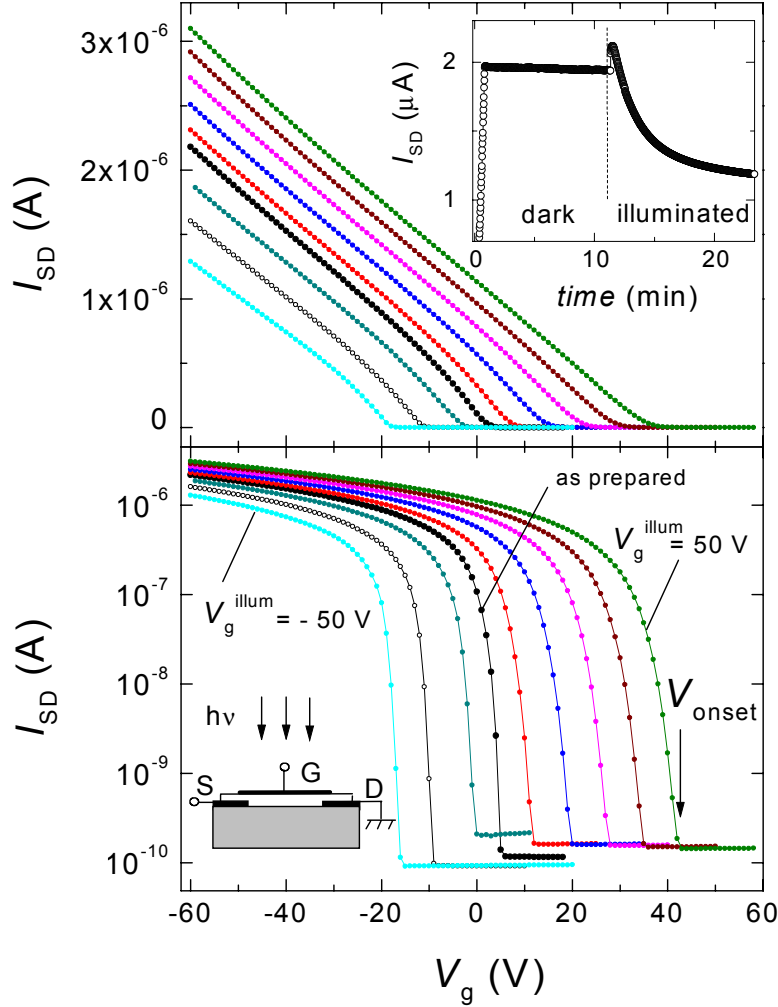

**Figure 1.** Linear (top) and semi-log (bottom) plots of the trans-conductance characteristics $I_{SD}(V_g)$ of a rubrene OFET, measured at a fixed $V_S = 5$ V in the dark after illumination of the device with a white light (intensity $P = 25$ mW/cm$^2$). During the illumination, different $V_g^{illum}$ (at $V_S = 0$) were applied: for the curves from right to left, $V_g^{illum} = 50, 40, 30, 20, 10, 0$ (as prepared device), $-10, -20, -50$ V. The illumination time $\Delta t$ was 2 min for positive $V_g^{illum}$ and 10-40 min for negative $V_g^{illum}$. After recording each curve, the initial state of the device (i. e., the "as prepared" curve) was restored by illumination at $V_g^{illum} = 0$ for 5 min. The top inset shows that $I_{SD}$, measured at fixed $V_S = 5$ V and $V_g = -60$ V, rapidly decreases under illumination due to the $V_{onset}$ shift. The bottom inset shows schematically the device geometry.

Regardless of the initial conditions, the magnitude of $V_{onset}$ after sufficiently long illumination depends only on the magnitude of $V_g^{illum}$. The rate of $V_{onset}$ shift, however, depends on several parameters, such as the intensity and wavelength of light and the transverse electric field at the semiconductor-dielectric interface (see below). The inset in the top panel of Fig. 1 shows that illumination of an operating device ($V_g = -60$ V) results in a rapid decrease of the conductivity of the channel, associated with the photo-induced shift of $V_{onset}$. Note an unusual sign of the effect: the current is *decreased* by illumination, contrary to what is usually observed in experiments on photoconductivity. The rate of the change of $I_{SD}$ under illumination varies with time. The photo-induced $V_{onset}$ shift is preserved in the dark for days regardless of further measurements at different $V_g$; the field-effect onset can be changed only by illumination at a different $V_g^{illum}$.

We believe that the observed photo-induced shift of $V_{onset}$ is caused by charging of the gate dielectric with the non-equilibrium carriers, photo-generated in the rubrene near the interface and transferred into the polymer. Depending on the sign of the transverse electric field at the rubrene/parylene interface during illumination, either positive or negative charges are transferred and trapped in parylene. For brevity, below we will refer to the positive and negative charges as "holes" and "electrons", respectively. However, we note that the nature of these non-equilibrium carriers and, in particular, the role of polaronic effects in the process of their generation requires further studies. In this model, variation of the surface density of charges transferred and immobilized in the polymer, $\Delta n$, is related to the observed shift of the onset voltage, $\Delta V_{onset}$, as $\mathbf{e}\Delta n = C_i \cdot \Delta V_{onset}$, where $\mathbf{e}$ is the elementary charge. Correspondingly, the transverse electric field, which induces the mobile polaronic charges at the semiconductor-dielectric interface, can be expressed as $E_\perp = (V_g - V_{onset})/d = V_g/d - \mathbf{e}n/(\varepsilon\varepsilon_0)$, where $d$ and $\varepsilon$ are the thickness and dielectric constant of the gate dielectric, respectively.

It is worth noting that the charge transfer into parylene affects not only the field-effect onset but also (to a much smaller extent) the mobility of field-induced polarons in the conduction channel, $\mu$, which is proportional to the slope of trans-conductance characteristics. In rubrene devices the "gating-under-illumination" procedure results in a small decrease of $\mu$, which saturates after multiple positive and negative shifts of the field-effect threshold. In tetracene OFETs, however, the decrease of $\mu$ with $V_{onset}$ shifts is more pronounced. The decrease of the mobility might indicate an additional contribution to the polaronic scattering or trapping caused by spatial fluctuations of the electrostatic field at the rubrene-parylene interface, created by the charges trapped in the dielectric.

The charge-transfer rate can be determined from the shift $\Delta V_{onset}$ after illumination for a (short) time interval $\Delta t$: $dn/dt = (C_i/e)(\Delta V_{onset}/\Delta t)$. Alternatively, it can be extracted from the time evolution of the field induced current $I_{SD}(t)$ measured at fixed $V_S$ and $V_g < V_{onset}$ under illumination (see the inset in Fig. 1): $dn/dt = (1/e\mu)(L/(WV_S))(dI_{SD}/dt)$. Both methods yield consistent results for the charge transfer rates.

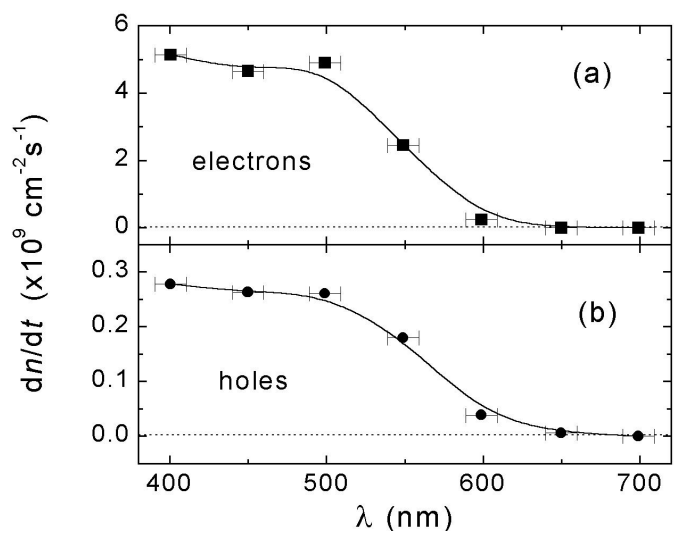

**Figure 2**. Spectral dependence of the rates of electron (a) and hole (b) transfer from rubrene to parylene. The photon flux is the same for all $\lambda$ ($\Phi = 5*10^{14}$ photons/cm$^2$s). The horizontal error bars represent the bandwidth of the interference filters ($\Delta\lambda = 20$ nm); the solid lines are the guide to the eye. In panel (**a**), $dn/dt$ was measured at $V_g^{illum} = +50$ V applied during the illumination; the initial zero-threshold state of the device has been prepared before the measurement at each $\lambda$ by illumination at $V_g = V_{SD} = 0$ with white light for 10 min. In panel (**b**), $dn/dt$ was measured at $V_g^{illum} = -60$ V; at each $\lambda$, the initial $V_{onset}$ was $+38$ V.

To characterize the spectral response of the observed effect, we have measured the rate of photo-induced charge transfer, $dn/dt$, as a function of the light wavelength ($\lambda$) at a fixed photon flux $\Phi = P/h\nu$ (P and $\nu$ are the light intensity and frequency, respectively) (Fig. 2). In both cases, the "red boundary" of the effect (steep increase of $dn/dt$ with decreasing $\lambda$) is observed close to the absorption edge in rubrene: $\lambda^0 = 560$ nm [16].

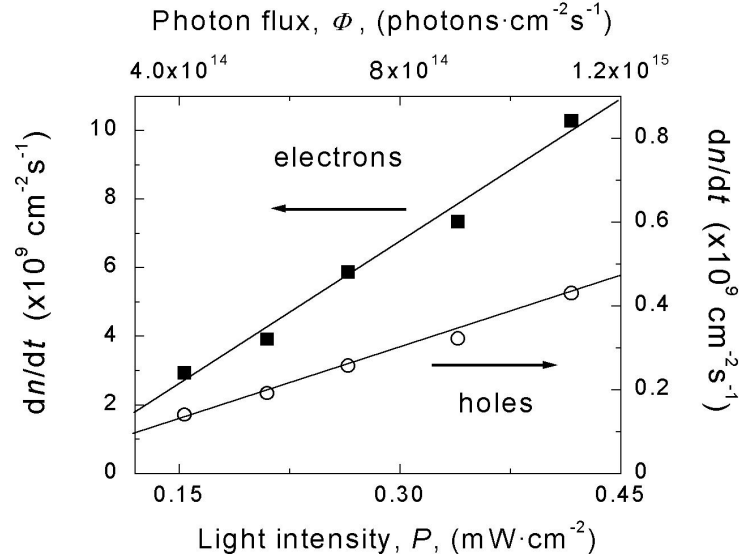

**Figure 3**. The charge transfer rate, $dn/dt$, for electrons (squares) and holes (circles) as a function of the photon flux $\Phi$ ($\lambda$ = 500 nm). At each light intensity, the same initial zero-threshold state of the device has been prepared by illumination at $V_g = V_{SD} = 0$ with white light for 10 min. The rates of electron and hole transfer have been measured at similar magnitudes of $E_\perp$ at the interface: $V_g^{illum} = +50$ V for the electron transfer and $V_g^{illum} = -60$ V for holes.

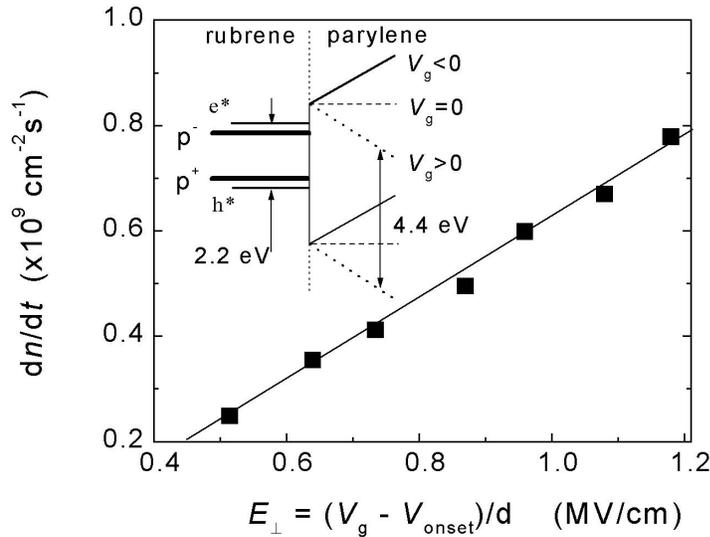

**Figure 4**. The dependence of the hole transfer rate on the transverse electric field, $E_\perp$, at the rubrene-parylene interface ($\lambda$ = 500 nm, photon flux $\Phi = 5 \cdot 10^{14}$ cm$^{-2}$s$^{-1}$). Before each measurement, the initial state of the device with the same $V_{onset} = +35$ V was prepared by illumination at $V_g^{illum} = +50$ V and $V_S = 0$. The inset shows schematically the energy diagram at the rubrene-parylene interface. The relaxed polaronic states p$^+$ and p$^-$ in rubrene occupy the energy levels within the HOMO-LUMO gap (2.2 eV). In parylene, the HOMO-LUMO gap is 4.4 eV.

Figure 3 shows that the charge transfer rate for holes and electrons from rubrene into parylene is proportional to the photon flux. Note that the rate of electron transfer is by a factor of ~ 30 higher than that for holes. The transfer rates for both processes depend linearly on the transverse electric field at the rubrene-parylene interface (for holes, this dependence is illustrated in Fig. 4). In the expression for the total transverse electric field at the interface, $E_\perp = (V_g - V_{onset})/d$, the absolute value of $V_{onset}$ is proportional to the charge trapped in the dielectric and increases with illumination time. This results in screening of $V_g$ and a decrease of $|E_\perp|$ during the illumination. For this reason, the charge transfer rate $dn/dt \propto - dI_{SD}/dt$ decreases with time (see inset in Fig. 1) [17]. We have measured $dn/dt$ as a function of $V_g^{illum}$ at a fixed initial $V_{onset}$ set prior each measurement, and as a function of the initial $V_{onset}$ at a fixed $V_g^{illum}$. In both cases, $dn/dt$ was found to be linear with $V_g$ and $V_{onset}$ (Fig. 4).

The experimental results shown in Figs. 1-4 can be described by the following expression for the rate of charge transfer across the OMC/polymer interface:

$$dn/dt = \chi(\lambda) \cdot \Phi \cdot E_\perp(t),$$

where $E_\perp(t) = V_g/d - \mathbf{e}n(t)/\varepsilon\varepsilon_0$. The prefactor $\chi(\lambda)$ is a product of probabilities of (a) photo-excitation of electrons and holes in rubrene close to the interface and (b) their transfer into parylene. The value $\chi(\lambda)$ approaches zero for the photon energy smaller than the optical HOMO-LUMO gap in rubrene: only non-equilibrium "hot" carriers contribute to the charge transfer. By contrast, the relaxed polarons, responsible for the charge transport in the conduction channel, remain in rubrene regardless of the magnitude of $E_\perp$. This is evidenced by the absence of the bias stress effect in rubrene OFETs. Thus, for an efficient charge transfer, the non-equilibrium carriers must be generated in rubrene close to the interface, within the length of thermalization of hot carriers $L_R$. For typical OMCs, $L_R$ does not exceed 5 nm [1], which indicates that only two-three molecular layers adjacent to the interface are relevant to the hot-carrier generation and transfer process. According to the Bouguer-Lambert law of exponential decay of light intensity in a medium, only a small fraction (~$10^{-3}$) of all incoming photons with energies greater than the HOMO-LUMO gap is absorbed within $L_R$ near the interface (typically, the absorption length in OMC is ~ 100 nm). For a photon flux $\Phi = 5 \cdot 10^{14}$ cm$^{-2}$s$^{-1}$, approximately $5 \cdot 10^{11}$ cm$^{-2}$s$^{-1}$ photons are absorbed within $L_R$ from the interface. Comparison of this quantity with the measured charge transfer rate $dn/dt \sim 5 \cdot 10^9$ cm$^{-2}$s$^{-1}$ for electrons and $\sim 3 \cdot 10^8$ cm$^{-2}$s$^{-1}$ for holes provides an estimate for the quantum efficiency of the hot carrier generation and their transfer into parylene, $\eta \equiv (dn/dt)/\Phi$: $\eta_e \sim 10^{-2}$ for electrons and $\eta_h \sim 0.6 \cdot 10^{-3}$ for holes.

Qualitatively the observed experimental facts can be interpreted on the basis of a schematic energy diagram shown in the inset in Fig. 4. Light absorption in rubrene leads to generation of "hot" carriers at the HOMO and LUMO levels. Clarification of the details of this process in rubrene requires further studies; recent optical experiments with ultra-short optical pulses favor direct generation of electrons and holes as primary photo-excitations versus excitons undergoing further dissociation [16,18]. Typically, the energies of "hot" carriers at the HOMO (LUMO) levels exceed the energy of relaxed polaronic states in OMCs by 0.1-0.2 eV [1]. This energy difference seems to be crucial for the charge transfer across the interface. The optical HOMO-LUMO gap in parylene (4.4 eV) is much larger than that of rubrene (2.2 eV). Since the measured efficiency of the electron transfer is greater than that for holes, we speculate that the energy difference between the LUMO levels in rubrene and parylene is smaller than that between the HOMO levels. The applied transverse electric field facilitates the charge transfer by reducing the effective thickness of the potential barrier at the interface.

To summarize, we have observed the photo-induced charge transfer across the interface between an ordered small-molecule organic semiconductor and a polymer dielectric in the field effect experiments. Two conditions are required for the charge transfer: photo-excitation with the photon energy greater than the HOMO-LUMO gap of the semiconductor and a transverse electric field applied at the semiconductor-dielectric interface. Trapping of the transferred charge in the polymer results in a controllable and stable-in-the-dark shift of the field-effect threshold. This study contributes to understanding of light-induced processes in organic semiconductors and opens new possibilities for the band-gap engineering at organic interfaces. The observed effect might have important implications in organic electronics. It points to a new aspect of stability of organic devices, and might be important for practical applications due to the possibility of "programming" of the OFETs characteristics and patterning of the conduction channel with light.

We acknowledge V. M. Agranovich, I. Biaggio and O. Ostroverkhova for helpful discussions. This work has been supported by the NSF grants DMR-0405208 and ECS-0437932.

10